\documentclass[algorithms,article,accept,moreauthors,pdftex,10pt,a4paper]{mdpi} 
\firstpage{1} 
\articlenumber{x}
\doinum{10.3390/------}
\pubvolume{xx}
\pubyear{2017}
\copyrightyear{2017}
\history{}
\pdfoutput=1
\fancypagestyle{plain}{}
\makeatletter
\renewcommand{\@maketitle}{
	\ifthenelse{\equal{\@arttype}{Supfile}}{%
	\begin{flushleft}
		\fontsize{18}{18}\selectfont
		\raggedright
		\noindent\textbf{Supplementary Materials: \@Title}%
		\par
		\vspace{12pt}
		\fontsize{10}{10}\selectfont
		\noindent\boldmath\bfseries{\@Author}
	\end{flushleft}
		}{%
		\begin{flushleft}
		\ifthenelse{\equal{\@arttype}{Book}}{%
			}{%
			\vspace*{-1.75cm}
			}
		{
		\ifthenelse{\equal{\@arttype}{Preprints} \OR \equal{\@arttype}{Book}}{%
			}{%
			\ifthenelse{\equal{\@arttype}{Conference Proceedings Paper}}{%
				\includegraphics[height=1.2cm]{logo-conference}%
				\hfill \href{http://www.mdpi.com}{\includegraphics[height=1cm]{logo-preforum}}%
				}{%
				\ifthenelse{\equal{\@status}{submit}}{%
					\hfill \href{http://www.mdpi.com}{%
					\includegraphics[height=1cm]{logo-mdpi}}\vspace{0.5cm}%
					}{
					\href{http://www.mdpi.com/journal/\@journal}{
                    }%
					\hfill \ifthenelse{\equal{\@journal}{scipharm}}{%
						\href{http://www.mdpi.com}{\includegraphics[height=1cm]{logo-mdpi-scipharm}}%
						}{%
						}%
					}%
				}%
			}%
		\par
		}
		{
    		\vspace{14pt}
    		\fontsize{10}{10}\selectfont
		\ifthenelse{\equal{\@arttype}{Book}}{
			}{
    			\ifthenelse{\equal{\@arttype}{Preprints}}{
				\textit{Article}%
				}{%
				\ifthenelse{\equal{\@arttype}{Reprint}}{%
					\textit{\ifthenelse{\equal{\@originalarttype}{\@empty}}{Article}{\@originalarttype}}%
					}{%
					}%
				}
			}%
 	   	\par%
    		}
    		{
   	 	\vspace{-1pt}
  	  	\fontsize{18}{18}\selectfont
   	 	\boldmath\bfseries{\@Title}
   	 	\par
   	 	\vspace{15pt}
   	 	}
   		{
    		\boldmath\bfseries{\@Author}
    		\par
    		\vspace{-4pt}
    		}
    		\end{flushleft}
  	  }%
	}
\makeatother
\renewcommand{\cright}{}

\newlength{\normalparindent}
\AtBeginDocument{\setlength{\normalparindent}{\parindent}}

\usepackage{siunitx}
\usepackage{amssymb}
\usepackage[a]{esvect}
\usepackage{subfig}
\usepackage{float}
\usepackage{bm}
\usepackage{listings}
\usepackage{dcolumn}
\newcolumntype{d}[1]{D{.}{\cdot}{#1} }
\newcolumntype{Y}{D{.}{.}{1.2}}

\usepackage{algorithm}
\usepackage[noend]{algpseudocode}
\algnewcommand{\algorithmicand}{\textbf{ and }}
\algnewcommand{\algorithmicor}{\textbf{ or }}
\algnewcommand{\OR}{\algorithmicor}
\algnewcommand{\AND}{\algorithmicand}
\algnewcommand{\var}{\texttt}
\usepackage{setspace}

\algrenewcommand\alglinenumber[1]{{\sffamily\footnotesize#1}}
\makeatletter
\algrenewcommand\ALG@beginalgorithmic{\fontsize{10}{16}\selectfont}
\makeatother

\makeatletter
\newcommand{\algmargin}{\the\ALG@thistlm}
\makeatother
\newlength{\whilewidth}
\algdef{SE}[parWHILE]{parWhile}{EndparWhile}[1]
  {\parbox[t]{\dimexpr\linewidth-\algmargin}{%
     \hangindent\whilewidth\strut\algorithmicwhile\ #1\ \algorithmicdo\strut}}{\algorithmicend\ \algorithmicwhile}%
\algnewcommand{\parState}[1]{\State%
  \parbox[t]{\dimexpr\linewidth-\algmargin}{\linespread{0.7}\selectfont\strut #1\strut}}

\usepackage{environ}
\NewEnviron{myequation}{%
\begin{equation}
\scalebox{0.9}{$\BODY$}
\end{equation}}

\Title{Different approaches to community detection}

\Author{Martin Rosvall$^{1,*}$, Jean-Charles Delvenne$^{2}$, Michael T. Schaub$^{3,4}$, and Renaud Lambiotte$^{5}$}

\AuthorNames{Martin Rosvall, Jean-Charles Delvenne, Michael T. Schaub, and Renaud Lambiotte}

\address{\noindent
$^{1}$Integrated Science Lab, Department of Physics, Umeå University, Umeå, Sweden\\
$^{2}$ICTEAM, Louvain-la-Neuve, Belgium\\
$^{3}$Institute for Data, Systems, and Society, Massachusetts Institute of Technology, Cambridge, USA\\
$^{4}$Department of Engineering Science, University of Oxford, UK\\
$^{5}$Mathematical Institute,  University of Oxford, Oxford, UK}

\corres{\hangafter=1 \hangindent=1.05em \hspace{-1.5em} Correspondence: martin.rosvall@umu.se}

\abstract{%
A precise definition of what constitutes a community in networks has remained elusive. Consequently, network scientists have compared community detection algorithms on benchmark networks with a particular form of community structure and classified them based on the mathematical techniques they employ. 
However, this comparison can be misleading because apparent similarities in their mathematical machinery can disguise different reasons for why we would want to employ community detection in the first place.
Here we provide a focused review of these different motivations that underpin community detection. 
This problem-driven classification is useful in applied network science, where it is important to select an appropriate algorithm for the given purpose.
Moreover, highlighting the different approaches to community detection also delineates the many lines of research and points out open directions and avenues for future research.
}

\newcommand{\Network}{\mathcal{N}}
\newcommand{\Nodes}{\mathcal{V}}

\newcommand{\cling}{\mathbf{C}}

\renewcommand{\citep}{\cite}
\renewcommand{\href}[2]{\texttt{#2}}
\begin{document}

\begin{quote}
This chapter is an extended  version of \emph{The many facets of community detection in complex networks}, Appl.~Netw.~Sci.~2:~4~(2017) by the same authors.
\end{quote}




\section{Introduction}

A precise definition of what constitutes a community in networks has remained elusive. Consequently, network scientists have compared community detection algorithms on benchmark networks with a particular form of community structure and classified them based on the mathematical techniques they employ. 
However, this comparison can be misleading because apparent similarities in their mathematical machinery can disguise different reasons for why we would want to employ community detection in the first place.
Here we provide a focused review of these different motivations that underpin community detection. 
This problem-driven classification is useful in applied network science, where it is important to select an appropriate algorithm for the given purpose.
Moreover, highlighting the different approaches to community detection also delineates the many lines of research and points out open directions and avenues for future research.

While research related to community detection dates back to the 70s in mathematical sociology and circuit design~\cite{lorrain1971structural,Donath1972}, Newman's and Girvan's work on modularity in complex systems just over ten years ago revitalized the field of community detection, making it one of the main pillars of network science research~\cite{Newman2004,Newman2006}.
The promise of community detection, that we can gain a deeper understanding of a system by discerning important structural patterns within a network, has spurred a huge number of studies in network science.
However, it has become abundantly clear by now that this problem has no canonical solution.
In fact, even a general definition of what constitutes a community is still lacking.
The reasons for this are not only grounded in the computational difficulties of tackling community detection. 
Rather, various research areas view community detection from different perspectives, illustrated by the lack of a consistent terminology:
`network clustering', `graph partitioning', `community', `block' or `module detection' all carry slightly different connotations.
This jargon barrier creates confusion, as readers and authors have different preconceptions and intuitive notions are not made explicit.

We argue that community detection should not be considered as a well-defined problem, but rather as an umbrella term with many facets.
These facets emerge from different goals and motivations for what it is about the network that we want to understand or achieve, and lead to different perspectives on how to formulate the problem of community detection.
It is critically important to be aware of these underlying motivations when selecting and comparing community detection methods.
Thus, rather than an in-depth discussion of the technical details of different algorithmic implementations~\cite{Schaeffer2007,Fortunato2010,Coscia2011,Parthasarathy2011,Newman2012,Malliaros2013,Xie2013,Fortunato2016}, here we focus on the conceptual differences between different perspectives on community detection. 

By providing a problem-driven classification, however, we do \emph{not} argue that the different perspectives are unrelated.
In fact, in some situations, different mathematical problem formulations can lead to similar algorithms and methods, and the different perspectives can offer valuable insights.
For example, for undirected networks, optimizing the objective function modularity~\cite{Newman2004}, initially proposed from a clustering perspective, can be interpreted as optimizing both a particular stochastic block model~\cite{newman2016equivalence} and an auto-correlation measure of a particular diffusion process on the networks~\cite{Delvenne2013}.
In other situations, however, such relationships disappear. 

While some perspectives arguably are more principled than others, we do not assert that there is a particular perspective that is a priori better suited for any given network.
In fact, as in data clustering~\cite{guyon2009clustering}, no one method can consistently perform the best on all kinds of networks~\cite{peel2017ground}.
Community detection is an unsupervised learning task that is blind to a researcher's intent with the analysis.
Accordingly, to understand a particular method's usefulness, we must take the researcher's interest in the communities into context~\cite{VonLuxburg2012}.

\begin{figure*}[htb]
    \centering
    \includegraphics[width=\textwidth]{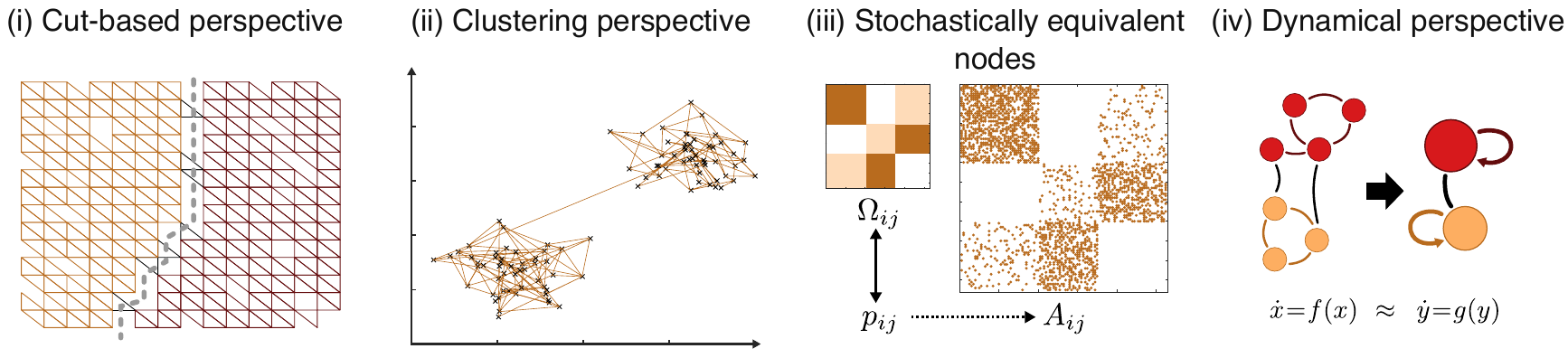}
    \caption{\textbf{Schematic of four different approaches to community detection.}  (\emph{i}) The cut-based perspective aims at minimising the number of links between groups of nodes, independently of their intrinsic structure. (\emph{ii}) The clustering perspective produces groups of densely connected nodes. (\emph{iii}) The stochastic equivalence perspective looks for groups in which nodes are stochastically equivalent, typically inferred through a generative statistical network model. (\emph{iv}) The dynamical perspective focuses on the impact of communities on dynamical processes and searches for dynamically relevant coarse-grained descriptions.}
    \label{fig:schematics}
\end{figure*}

In the following, we unfold different aims underpinning community detection---in a relaxed form that includes assortative as well as disassortative group structures with dense and sparse internal connections, respectively---and discuss how the resulting problem perspectives relate to various applications.
We focus on four broad perspectives that have served as motivation for community detection in the literature:
(\emph{i})~the cut-based perspective minimizes a constraint such as the number of links between groups of nodes; (\emph{ii}) the clustering perspective maximizes internal density in groups of nodes; (\emph{iii}) the stochastic block model perspective identifies groups of nodes in which nodes are stochastically equivalent; and (\emph{iv}) the dynamical perspective identifies groups of nodes in which flows stay for a relatively long time such that they form building blocks of dynamics on networks (see Fig.\ref{fig:schematics}).
While this categorization is not unique, we believe that it can help clarify concepts about community detection and serve as a guide to determining the appropriate method for a particular purpose.

\section{Minimizing constraint violations: the cut-based perspective}\label{sec:II}
An early network partitioning application was circuit layout and design~\cite{Alpert1995,Fortunato2010}.
This application spurred development of the now classical Kernighan-Lin algorithm~\cite{Kernighan1970} and the work by Donath and Hoffmann~\cite{Donath1972,Donath1973}, who were among the first to suggest the use of eigenvectors for network partitioning.
For example, we might be confronted with a network that describes the signal flows between different components of a circuit.
To design the circuit in an efficient way, our goal is now to partition the network into a fixed number of approximately equally sized groups for balanced load with a small number of edges between those groups for minimal communication overhead.
The edges that run between the groups are commonly denoted as the cut.
To design the most efficient circuit, our aim is thus to minimise this cut with more or less balanced groups.

To make this more precise, let us consider one specific variant of this scheme, known as \textit{ratio cut}~\cite{Hagen1992}.
Let us denote the adjacency matrix of an undirected network $\Network$ with $n$ nodes by $A$, where $A_{uv} = 1$ if there is a connection from node $u$ to node $v$, and $A_{uv} =0$ if there is no connection. 
We can now write the problem of optimizing the ratio cut for a bipartition of all nodes $\Nodes$ into two communities $\Nodes_1$ and $\Nodes_2 = \mathcal{V}\backslash\mathcal{V}_1$ as follows~\cite{Hagen1992,VonLuxburg2007}:
\begin{flalign}
    \min_{\Nodes_1} \text{RatioCut}(\Nodes_1, \Nodes_2) := \min_{\Nodes_u} \sum_u \dfrac{\text{cut}(\Nodes_u,\Nodes \backslash \Nodes_u)}{|\Nodes_u|},
\end{flalign}
where $\text{cut}(\mathcal{V}_1,\mathcal{V}_2) := \sum_{u\in\mathcal{V}_1, v \in \mathcal{V}_2} (A_{uv} + A_{vu})/2$ is the sum of the possibly weighted edges between the two vertex sets $\Nodes_1, \Nodes_2$.
Related problem formulations also occur in the context of parallel computations and load scheduling~\cite{Spielman1996,Pothen1997}, where approximately equally sized portions of work are to be sent to different processors, while keeping the dependencies between those tasks minimal.
Further applications include scientific computing~\cite{Spielman1996,Pothen1997}, where partitioning algorithms divide the coordinate meshes when discretising and solving partial differential equations.
Image segmentation problems may also be phrased in terms of cut-based measures~\cite{Shi2000,VonLuxburg2007}. 

Investigating these types of problems has led to many important contributions to partitioning networks, in particular in relation to spectral methods.
The connection between spectral algorithms and cut-based problem formulations arises naturally by considering relaxations of the original, combinatorially hard discrete optimisation problems, such as Eq.~\eqref{eq:ratio_cut}, or other related objective functions such as the average or normalised cuts.
This can be best seen when rewriting the above optimisation problem as follows:
\begin{flalign}\label{eq:ratio_cut}
    \min_f & \quad f^TLf \\
    \text{subject to} & \quad f \perp \mathbf{1} \quad \|f\|= \sqrt{n}\\
    \text{where} & \quad f_u:= 
    \begin{cases}
        -\sqrt{|\Nodes_2|/|\Nodes_1|} \quad \text{if } u\in \Nodes_1\\
        \;\;\;\sqrt{|\Nodes_1|/|\Nodes_2|} \quad \text{if } u\in \Nodes_2
    \end{cases}
\end{flalign}
Here the Laplacian matrix of the network has been defined as $L=D-A$, where $D$ is the diagonal degree matrix with $D_{uu} = \sum_{v}A_{uv}$.
Fiedler realised already in the 70s that the second smallest eigenvalue of the Laplacian matrix is associated with the connectivity of the network, and that the associated eigenvector thus can be used to compute spectral bi-partitions~\cite{Fiedler1973,Fiedler1975}.
Such spectral ideas led to many influential algorithms and methods; see, for example, von Luxburg~\cite{VonLuxburg2007} for a tutorial on spectral algorithms.

In this cut-based problem formulation, there is no specification as to how the identified groups in the partition should be connected internally.  While the implicit constraint is that the groups must not split into groups with an even smaller cut, there is no specification that the groups of nodes should be densely connected internally.
Indeed, the type of networks considered in the context of cut-based partitions are often of a mesh- or grid-like form, for which several guarantees can be given in terms of the quality of the partitions obtained by spectral algorithms~\cite{Spielman1996}.
While such non-dense groupings emerging from the analysis of non-clique structures~\cite{Schaub2012} can also be dynamically relevant (see section \ref{sec:dynamics}), they are likely missed when employing a community notion that focuses on finding dense groupings, as discussed next.

\section{Maximizing internal density: the clustering perspective}
\label{sec:clustering}
A different motivation for community detection arises in the context of data clustering.
We use the term clustering, which can have many definitions, in the following sense:
For a set of given data points in a possibly high-dimensional space, the goal is to partition the points into a number of groups such that points within a group are close to or similar to each other in some sense, and points in different groups are more distant from each other.
To achieve this goal, one often constructs a proximity or similarity network between the points and tries to group together nodes that are closer to each other than they are to the rest of the network.
This approach results in a form of community detection problem where the closeness between nodes is described by the presence and weight of the edges between them.

Although minimizing the cut size and maximizing the internal number of links are closely related, there are differences pertaining to the typical constraints and search space associated with these objective functions.
First, when employing a clustering perspective, there is normally no \textit{a priori} information about the number of groups we are looking for.
Second, we do not necessarily require the groups to be balanced in any way; rather we would like to find an optimal split into densely knit groups irrespective of their relative sizes.

Unsurprisingly, finding an optimal clustering is a computationally difficult problem.
Further, as Kleinberg has shown~\cite{Kleinberg2003}, there are no clustering algorithms that satisfy a certain set of intuitive properties we might require from a clustering algorithm in continuous spaces. Similar problems also arise in the discrete setting for clustering of networks~\cite{browet2017incompatibility}.

Nevertheless, there exists a large number of methods that follow a clustering-like paradigm and separate the nodes of a network into cohesive groups of nodes, often by optimizing a quality function.
An important clustering metric in this context is the so-called conductance~\cite{Kannan2004,Andersen2006,Spielman2013,Kloster2014}.
Optimizing the global conductance was introduced as a way to produce a global bi-partition similarly to the 2-way ratio-cut.
However, this quantity has been successfully employed more recently as a local quality function to find localised clusters around one or more seed nodes.
The local conductance of a set of nodes $\Nodes_q\subset \Nodes$ can be written
\begin{equation}
    \phi(\Nodes_q) := \frac{\sum_{u\in \Nodes_q, v \notin \Nodes_q}A_{uv}}{\min\{\text{vol}(\Nodes_q),\text{vol}(\Nodes - \Nodes_q )\}},
\end{equation}
where $\text{vol}(\Nodes_q) := \sum_{u\in \Nodes_q}\sum_v A_{uv}$ is the total degree of the nodes in set $\Nodes_q$, commonly called its volume in analogy with geometric objects.
Interestingly, it has been shown that, in specific contexts, the conductance can be a good predictor of some latent group structures in real-world applications~\cite{Yang2015}. 

Moreover, a local perspective on community detection has two appealing  properties: 
First, the definition of a cluster does not depend on the global network structure but only on the relative local density.
Second, only a portion of a network needs to be accessed, which is advantageous if there are computational constraints in using large networks, or we are only interested in a particular subsystem.
In such cases, we would like to avoid having to apply a method to the whole network in order to find, for example, the cluster containing a particular node in the network.

The Newman-Girvan modularity~\cite{Newman2004,Newman2006} is arguably one of the most common clustering measures used in the literature and was originally proposed from the clustering perspective discussed here.
It is a global quality function and aims to find the community structure of the network as a whole.
Given a partition $\cling =\{\Nodes_1, \ldots, \Nodes_k\}$ of a network into $k$ groups, the modularity of $\cling$ can be written as:
\begin{equation}
    Q(\cling) := \dfrac{1}{2m}\sum_{q =1}^k\sum_{u,v \in \Nodes_q} \left[A_{uv} - \dfrac{d_ud_v}{2m}\right],
\end{equation}
where $d_u = \sum_v A_{uv}$ is the degree of node $u$ and ${2m = \sum_u d_u}$ is the total weight of all edges in the network.
By optimizing the modularity measure over the space of all partitions, one aims to identify groups of nodes that are more densely connected to each other than one would expect from a statistical null model of the network.
This statistical null model is commonly chosen to be the configuration model with preserved degree sequence.

However, a by-product of this choice of a global null-model is the tendency of modularity to balance the size of the groups in terms of their total connectivity.
While different variants of modularity aim to account for this effect~\cite{Fortunato2010}, it means modularity can be interpreted as a trade-off between a cut-based measure and an entropy~\cite{Delvenne2013}.
Modularity is typically optimized with spectral or greedy algorithms~\cite{Fortunato2010,Newman2006a,Blondel2008}.
While there are problems with modularity, such as its resolution limit~\cite{Fortunato2007} and other spurious effects~\cite{Fortunato2007,Good2010,Guimera2004,Lancichinetti2011a}, the general idea has triggered researchers to develop a plethora of algorithms that follow a similar strategy~\cite{Fortunato2010}.
Several works have addressed some of the shortcomings, by incorporating a resolution parameter, for example, or by explicitly accounting for the density inside each group~\cite{Chen2014,Chen2015}. In practice, however, less seems to beat more and the original formulation of modularity remains the most widely used.

\section{Identifying structural equivalence: the stochastic block model perspective}
By grouping \textit{similar} nodes that link to similar nodes within communities, we constrain ourselves to finding \textit{assortative} group structure~\cite{Fortunato2016}.
While we may also have hierarchical clusters with clusters of clusters, etc., such an assortative structural organisation is too restrictive if we want to define groups based on more general connectivity patterns that include disassortative communities with weaker interactions within rather than between communities.

In social network analysis, a common goal is to identify nodes within a network that serve a similar structural role in terms of their connectivity profile.
Accordingly, nodes are similar if they share the same kind of connection patterns to other nodes~\cite{lorrain1971structural}.
This idea is captured in concepts such as \textit{regular equivalence}, which states that nodes are regularly equivalent if they are equally related to equivalent others~\cite{Everett1994,Hanneman2005}. The first algorithms for identifying groups of ``approximately equivalent'' nodes were deterministic and permuted adjacency matrices to reveal block structures in so-called block models~\cite{white1976social,arabie1978constructing}.

A relaxation of regular equivalence is \emph{stochastic equivalence}~\cite{Holland1983}, where nodes are equivalent if they connect to equivalent nodes with equal probability. The stochastic formulation generalises observations and forms generative models, which can be used for prediction. Because of this advantage over non-stochastic formulations, we focus on stochastic equivalence.

One of the most popular techniques to model and detect stochastically equivalent relationships in network data is to use stochastic block models (SBMs)~\cite{Holland1983,Nowicki2001} and associated inference techniques.
These models have their roots in the social networks literature~\cite{Holland1983,Anderson1992}, and provide a flexible framework for modelling block structures within a network.
When considering block models, we are interested in identifying node groups such that nodes within a community connect to nodes in other communities in an `equivalent way'~\cite{Fortunato2016}.

Consider a network composed of $n$ nodes divided into $k$ classes.
The standard SBM is defined by a set of node class labels and the affinity matrix $\Omega$.
More precisely, the link probability between two nodes $u, v$ belonging to class $c_u$ and $c_v$ is given by: 
$$p_{uv} := \mathbb P(A_{uv})= \Omega_{c_uc_v}.$$
Under an SBM, nodes within the same class share the same probability of connecting to nodes of another class. This is the mathematical formulation of having stochastically equivalent nodes within each class.
Finding the latent groups of nodes in a network now amounts to inferring the model parameters that provide the best fit for the observed network. That is, find the SBM with the highest likelihood of generating the data.

The standard SBM assumes that the expected degree of each node is a Poisson binomial random variable, a binomial random variable with possibly non-identical success probabilities in each trial.
Because inferring the most likely SBM typically results in grouping nodes based on their degree in empirical networks with broad degree distributions, it can be advantageous to include a degree-correction into the model. 
In the degree corrected SBM~\cite{Karrer2011}, the probability $p_{uv}$ that a link will appear between two nodes $u, v$ depends both on their class labels $c_u, c_v$ and their respective degree parameters $d_i, d_j$ (each entry $A_{ij}$ might be a Bernoulli or a Poisson random variable such as in~\cite{Karrer2011}):
$$p_{uv} \sim d_u d_v \Omega_{c_uc_v}.$$ 
Thus, while edges in real-world networks tend to be correlated with effects such as triadic closure~\cite{Fortunato2010}, by construction edges are conditionally independent random variables in SBMs. 
Moreover, most common SBMs are defined for unweighted networks or networks with integer weights by modelling the network as a multi-graph.
Though generalizations are available~\cite{Aicher2014,peixoto2015inferring}, this is still a less studied area.

In contrast to the notions of community considered above, with stochastic equivalence we are no longer interested in maximising some internal density or minimising a cut. 
To see this, consider a bipartite network that from a cut- or density-based perspective contains no communities.
From the stochastic equivalence perspective, however, we would say that this network contains two groups because nodes in each set only connect to nodes in the other set.
When adopting an SBM to detect such structural organisation of the links, we explicitly adopt a statistical model for the networks.
The network is essentially an instance of an ensemble of possible networks generated from such a model.\footnote{This ensemble assumption is also reflected in the modularity formalism, where the observed network is compared to a null model.}

This model-based approach comes with several advantages:
First, by defining the model, we effectively declare what is signal and what is noise in the data under the SBM.
We can thus provide a statistical assessment of the observed data with, for example, $p$-values under the SBM.
In other words, we can identify patterns that cannot be reasonably explained from density fluctuations of edges inherent to any realisation of the model.
Second, we are able, for example, to generate new networks from our model with a similar group structure, or predict missing edges and impute data.
Third, we can make strong statements about the detectability of groups within a network.
For example, precise criteria specify when any algorithm can recover the planted group structure for a network created by an SBM~\cite{Decelle2011a,Mossel2013}.
By fitting an SBM to an observed adjacency matrix, it is possible to recover such a planted group structure down to its theoretical limit~\cite{Mossel2013,Massoulie2014}.
These criteria apply to networks generated with SBMs and not real networks in general, in which case we do not know what kind of process created the network~\cite{peel2017ground}. It is nevertheless a remarkable result since it highlights the fact that there are networks with undetectable block patterns.

Moreover, this model-based approach also offers ways to estimate the number of communities from the data by some form of model selection, including hypothesis testing~\cite{Bickel2016}, spectral techniques~\cite{Krzakala2013,Saade2014}, the minimum description length principle~\cite{Peixoto2013}, or Bayesian inference~\cite{yan2016bayesian}.

Finally, the generative nature of SBMs also makes them well suited for constructing benchmark networks.
As a consequence, many benchmark networks proposed in the literature, such as the commonly used LFR benchmarks~\cite{Lancichinetti2008}, are specific types of SBMs.
Results on these benchmark networks should therefore be taken for what they are: the ability to recover the underlying group structure of specific types of SBM-generated networks. For example, sparse networks without any underlying group structure still can contain meaningful dynamical building blocks.

\section{Identifying coarse-grained descriptions: the dynamical perspective}\label{sec:dynamics}
Let us now consider a fourth alternative motivation for community detection, focusing on the processes that take place on the network. 
All notions of community outlined above are effectively structural in the sense that they are mainly concerned with the composition of the network itself or its representation as an adjacency matrix.
However, in many cases one of the main reasons to apply tools from network science is to understand the \textit{behaviour} of a system.
While the topology of a system puts constraints on the dynamics that can take place on the network, the network topology alone cannot explain the system behaviour.
For example, instead of finding a coarse-grained description of the adjacency matrix, we might be interested in finding a coarse grained description of the dynamics acting on top of the network with multi-step paths beyond the nearest neighbours.

Take air traffic as an example. 
An airline network, with weighted links connecting cities according to the number of flights between them, can offer some interesting insights about air traffic.
For instance, in the US air traffic network based on the number of flying passengers, Las Vegas and Atlanta form two major hubs. 
However, if we focus instead on the passenger flows based on actual multi-leg itineraries, the two cities show very different behaviours: Las Vegas is a tourist destination and typically the final destination of itineraries, whereas Atlanta is often a transfer hub to other final destinations~\cite{Rosvall2014,peixoto2017modelling}. 
Thus, these airports play dynamically quite different roles in the network.
Focusing on interconnection patterns alone can give an incomplete picture if we are interested in the dynamical behaviour of a system, for which additional dynamical information should be taken into account.
Conversely, a concentration of edges with high impact on the dynamics may arise just from a statistical fluctuation, if the network is seen as a realization of a particular random network model.
In this way, structural and dynamical approaches can offer complementing information. 

In general, however, they are blocks of nodes with different identities that trap the flow or channel it in specific directions. 
That is, they form reduced models of the dynamics where blocks of nodes are aggregated to single meta nodes with similar \textit{dynamical function} with respect to the rest of the network. 
In this view, the goal of community detection is to find effective coarse-grained system descriptions of how the dynamics take place on the network structure.

To induce multi-step paths and couple also non-neighbouring nodes, the dynamical approach to community detection has primarily focused on modelling the dynamics with Markovian diffusion processes~\cite{Rosvall2008,Delvenne2010,Lambiotte2014}, though the work of topological scales and synchronization share the same common ground~\cite{Arenas2006}.
Interestingly, for simple diffusion dynamics such as a random walk on an undirected network, which is essentially determined by the spectral properties of the network's Laplacian matrix, this perspective is tightly connected to the clustering perspective discussed in section \ref{sec:clustering}.
This is because the presence of densely knit groups within the network can introduce a time-scale separation in the diffusion dynamics:
A random walker traversing the network will initially be trapped for a significant time inside a community corresponding to the fast time-scale, before it can escape and explore the larger network corresponding to a slower time-scale.
However, this connection between link density and dynamical behaviour breaks down for directed networks, even for a simple diffusion process~\cite{Rosvall2008,Lambiotte2014,Schaub2012}.
This apparent relationship breaks down completely when focusing on longer pathways, possibly with memory effects in the dynamics~\cite{Rosvall2014,Salnikov2016}.

A dynamical perspective is useful especially in applications in which the network itself is well defined, but the emergent dynamics are hard to grasp.
For instance, consider the nervous system of the roundworm \textit{C. elegans}, for which there exists a distinct network.
A basic generative network model, such as a Barabasi-Albert network or an SBM, might be too simple to capture the complex architecture of the network, and sampling alternative networks from such a model will not create valid alternative roundworm connectomes. 
Indeed, some more complicated network generative models have been proposed to model the structure of the network~\cite{Nicosia2013}, and may be used to assess the significance of individual patterns compared to the background of the assumed model.
However, if we are interested instead in assessing the dynamical implications of the evolutionary conserved network structure, it may be fruitful to engineer differences in the actual network and investigate how they affect the dynamical flows in the system.
For instance, one can replicate experimental node ablations \textit{in silico} and assess their dynamical impact~\cite{Bacik2016}.

In the dynamical perspective, we are typically interested in how short-term dynamics integrate into long-term behaviour of the system and seek a coarse-grained description of the dynamics occurring on a given network.
That is, the network itself represents the true structure, save for empirical imperfections. 
Therefore, in the dynamical perspective, model selection is in general not about comparing competing models~\cite{Peixoto2013,yan2016bayesian} but about comparing coarse-grained descriptions of the dynamics on resampled realisations of the observed network with, for example, the bootstrap \cite{rosvall2010mapping} or cross-validation \cite{persson2016maps}. Nevertheless, it is possible to formulate generative statistical models for empirically observed pathways~\cite{peixoto2017modelling}. However, whereas the generative approach in, for example, ref.~\cite{peixoto2017modelling} explicitly models the underlying state space of trajectories, we may simply be interested in effectively compressing the long-term behaviour of the system~\cite{persson2016maps}.

Two methods that exploit the long-term dynamics of the system by identifying communities with long flow persistence are the Markov stability~\cite{Delvenne2013} and the map equation~\cite{Rosvall2008}. Whereas the Markov stability takes a statistical approach and favours communities inside which a random walker is more likely to remain after a given time $t$ than expected at infinite time, the map equation reveals modular regularities by compressing the dynamics. It is an information-theoretic approach that uses the duality between compressing data and finding regularities in the data~\cite{Rosvall2008,shannon1948mathematical}. It measures the quality of communities by how much they can compress a modular description of the dynamics. The shorter description, the more detected regularities, such that the shortest description captures the most regularities.  Given module assignments $\cling$ of all nodes in the network, the map equation measures the description length $L(\cling)$ of a random walker that moves within and between modules from node to node by following the links between the nodes~\cite{rosvall2009map}: 
\begin{align}\label{eq:mapequation}
L(\cling) = q_{\curvearrowleft} H(\mathcal{Q}) + \sum_{q=1}^{k}p_{\circlearrowright}^qH(\mathcal{P}^q)
\end{align}
Here the entropy $H(\mathcal{Q})$ measures the average per-step description length of movements between modules derived from module-enter rates $\mathcal{Q}$ of all $k$ modules and $H(\mathcal{P}^q)$ measures the average per-step description length of movements within module $q$ derived from node-visit and module-exit rates $\mathcal{P}^q$. The description lengths are weighted by their rate of use, $q_{\curvearrowleft}$ and $p_{\circlearrowright}^q$, respectively. The visit rates can be obtained by first calculating the PageRank of links and nodes or directly from the data if they represent flow themselves. In any case, finding the optimal partition of the network by assigning each node to one or more modules corresponds to testing different node assignments and picking the one that minimizes the map equation. This simple formulation allows for straightforward generalizations to coarse-grained hierarchical~\cite{rosvall2011multilevel} descriptions of dynamics in memory~\cite{Rosvall2014} and multilayer~\cite{de2015identifying} networks.

As the air traffic example above illustrates, it can be crucial to go beyond standard network abstractions and consider memory and higher-order effects in multi-step pathways to better understand system behaviour. For example, higher-order abstractions, such as memory and multilayer networks, provide principled means to reveal highly overlapping modular organization in complex systems: link clustering~\cite{Ahn2010} and clique percolation~\cite{palla2005uncovering} methods can be interpreted as trying to account for second-order Markov dynamics (see Supplementary Note 3 of ref.~\cite{Rosvall2014}).

Compared to some of the other perspectives, the dynamical viewpoint has received somewhat less attention and has been confined mainly to diffusion dynamics.
A key challenge is to extend this perspective to other types of dynamics and link it more formally to approaches of model order reduction considered in control theory.
In light of the recently growing interest in the control of complex systems, this could help us better understanding complex systems.

\section{Discussion}
Community detection can be viewed through a range of different lenses. 
Rather than looking at community detection as a generic tool that is supposed to work in a generic context, considering the application in mind is important when choosing between or comparing different methods.
Each of the perspectives outlined above has its own particularities, which may or may not be suitable for the problem of interest.

We emphasise the different perspectives in the following example.
Given a real-world network generated by a possibly complex random assignment of edges, we assume that we are interested in some particular dynamics taking place on this network, such as epidemic spreading.
We also assume that the network is structured such that the dynamics exhibit a time-scale separation.
If, for instance, we want to coarse-grain an epidemic and identify critical links that should be controlled to confine the epidemic, then it does not matter whether or not random fluctuations generated the modules that induce the time-scale separation. In any case, these modules will be relevant for the dynamics.

Assume now that the same network encodes interdependency of tasks in a load-scheduling problem.
In such a circumstance, a cut-based approach will find a relevant community structure, in that it will allow an optimally balanced assignment of tasks to processors that minimises communication between processors. 
These communities may be different from the ones attached to the epidemic-spreading example.

If we instead assume that the links represent friendships, we may want to identify densely knit groups irrespective of their relative sizes. Accordingly, taking the clustering perspective and maximizing the internal density can give yet another set of communities.

In these three cases, we considered a single realisation of the network with the goal of extracting useful information about its structure, independently of the possible mechanisms that generated it.

Let us finally consider the same network from a stochastic equivalence perspective, and assume for simplicity that the network is a particular realization of an Erd\H{o}s-R\'enyi network.
In this case, an approach based on the SBM is expected to declare that there is no significant pattern to be found here at all, as the encountered structural variations can already be explained by random fluctuations rather than by hidden class labels. 
Thus, communities in the SBM picture are defined via the latent variables within the statistical model of the network structure, and not via their impact on the behavior of the system.
In this way, different motivations for community detection can find different answers even for the very same network. 

\begin{figure}[htb]
    \centering
    \includegraphics[width=0.8\textwidth]{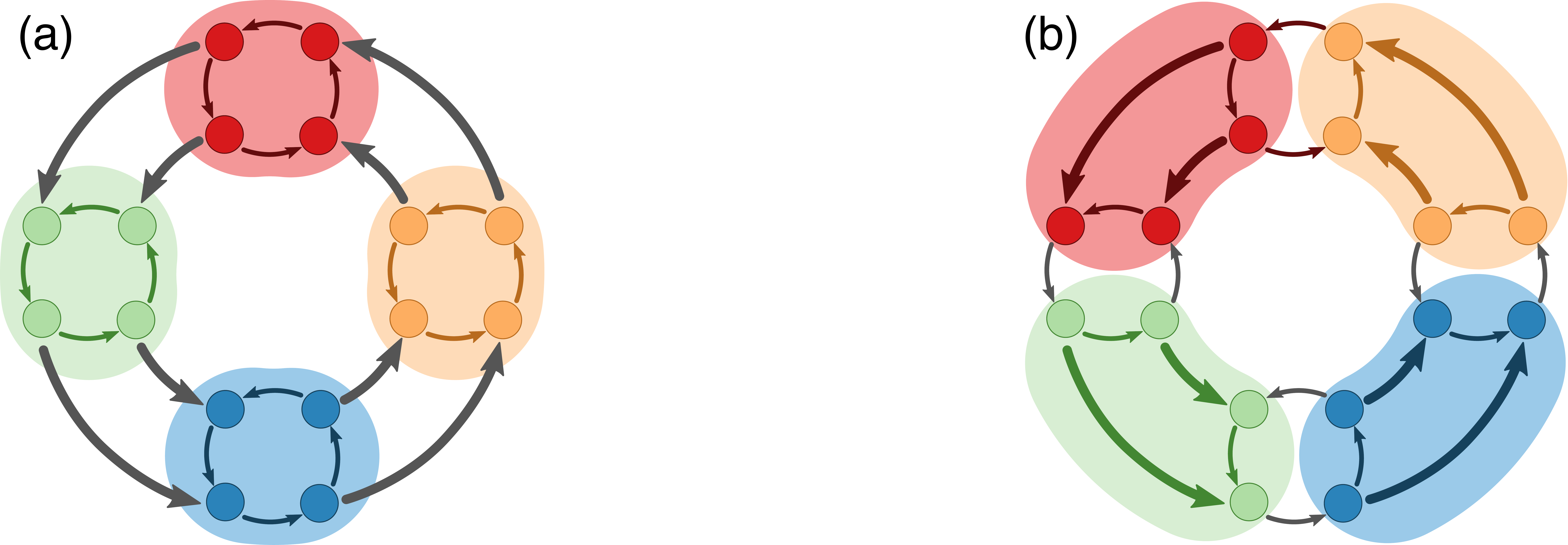}
    \put(-300,-20){Modularity}
    \put(-300,-20){\phantom{Map equation} $Q=0.25$}
    \put(-300,-32){Map equation $\bm{L=2.67}$ bits/step}
    \put(-115,-20){Modularity}
    \put(-115,-20){\phantom{Map equation} $\bm{Q=0.50}$}
    \put(-115,-32){Map equation $L=4.13$ bits/step}
    \caption{\textbf{Communities that highlight different aspects of networks.} Identifying coarse-graining flows in groups, here illustrated by the map equation, and densely connected groups, here illustrated by modularity, highlights different aspects of structure in directed and weighted networks. Each shaded area represents a cluster in two alternative clusterings of a schematic network. (a) The  clustering as optimised by the map equation (minimum $L$). (b) The clustering as optimised by modularity (maximum $Q$). The thicker links have double the weight of the thinner links. Example from ref.~\cite{Rosvall2008}.}
    \label{fig:networks}
\end{figure}
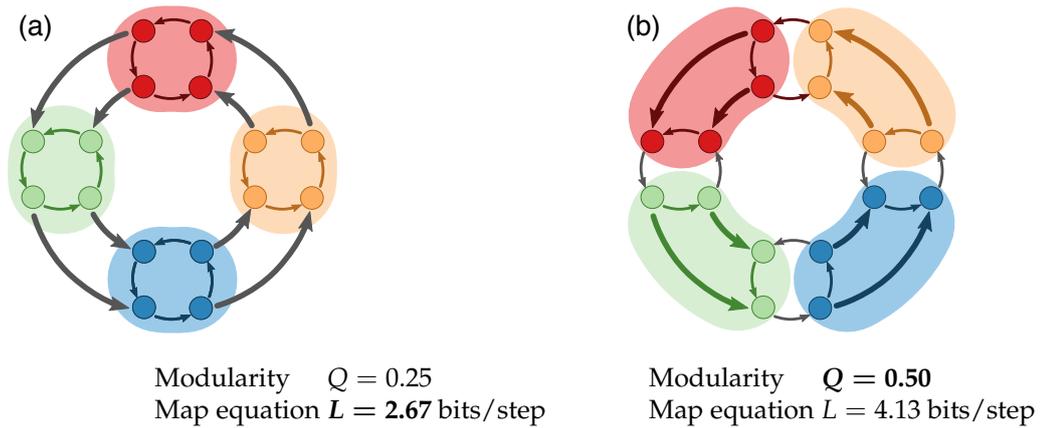

To illustrate that different motivations can give different answers for the same network, we use an example from ref.~\cite{Rosvall2008}. The directed, weighted network is formed as a ring of rings such that each internal ring captures flows for a relatively long time despite the stronger links between the rings (see Fig.~\ref{fig:networks}). For example, a random walker takes on average three steps within a ring highlighted as a cluster in Fig.~\ref{fig:networks}a before exiting. In contrast, a random walker takes on average only 2.4 steps within a cluster in Fig.~\ref{fig:networks}b. A method that seeks to coarse-grain the dynamics will therefore identify the flow modules in Fig.~\ref{fig:networks}a rather than the clusters with high internal density in Fig.~\ref{fig:networks}b. For example, the modular description quantified by the map equation is almost twice as efficient with the flow modules as it is with the clusters with high internal density. The opposite is true for a method that highlights structural regularity and high internal density: the modularity score is twice as large for the clustering in Fig.~\ref{fig:networks}b.
While this example only illustrates the fundamental difference between two methods applied to a schematic network, methods from different perspectives will give different answers for real networks as well \cite{hric2014community}.

In addition to the differences \textit{between} these perspectives, there are also variations \textit{within} each perspective.
For instance, distinct plausible generative models such as the standard SBM or the degree-corrected SBM will, for a given network, lead to different inferred community structure.
Similar variations exist in the dynamical paradigm as well: Distinct natural assumptions for the dynamics, such as dynamics with or without memory, uniform across nodes or edges, etc., applied to a given network will lead to different partitions.
Also different balancing criteria (see section \ref{sec:II}) or different concepts of high internal density (see section \ref{sec:clustering}) will be valid in different contexts. 

In fact, some of the internal variations make the perspectives overlap in particular scenarios.
For instance, one can compare all the algorithms on simple, undirected LFR benchmark networks~\cite{Lancichinetti2008}.
However, the LFR benchmark clearly imposes a density-based notion of communities.
Similarly, for simple undirected networks, optimizing modularity corresponds to the inference of a particular SBM~\cite{newman2016equivalence} or may be reinterpreted as a diffusion process on a network~\cite{Delvenne2013}.
Nevertheless, this overlap of concepts, typically present in unweighted, undirected networks, is only partial, and breaks down, for example, in directed networks, or for more complex dynamics.

\section{Conclusions}

In summary, no general purpose algorithm will ever serve all applications or data types~\cite{peel2017ground}, because each perspective emphasizes a particular core aspect: A cut-based method provides good separation of balanced groups, a clustering method provides strong cohesiveness of groups with high internal density, stochastic block models provide strong similarity of nodes inside a group in terms of their connectivity profiles, and methods that view communities as dynamical building blocks aim to provide node groups that influence or are influenced by some dynamics in the same way.
As more and more diverse types of data are collected, leading to ever more complex network structures, including directed~\cite{Malliaros2013}, temporal~\cite{Holme2012,Sekara2016}, multi-layer or multiplex networks~\cite{Boccaletti2014}, the differences between the perspectives presented here will become even more striking---the same network might have multiple valid partitions depending on the question about the network we are interested in.
We might moreover not only be interested in partitioning the nodes, but also in partitioning edges~\cite{Ahn2010}, or even motifs~\cite{Benson2016}.
Rather than striving to find a `best' community-detection algorithm for a better understanding of complex networks, we argue for a more careful treatment of what network aspects we seek to understand when applying community detection.

\section{Acknowledgements}

\noindent We thank Aaron Clauset, Leto Peel and Daniel Larremore for fruitful discussions. 
MR was supported by the Swedish Research Council grant 2016-00796.
MTS, JCD, and RL acknowledge support from: FRS-FNRS; the Belgian Network DYSCO (Dynamical Systems, Control and Optimisation) funded by the Interuniversity Attraction Poles Programme initiated by the Belgian State Science Policy Office; and the ARC (Action de Recherche Concerte) on Mining and Optimization of Big Data Models funded by the Wallonia-Brussels Federation. MTS received funding from the European Union’s Horizon 2020 research and innovation programme under the Marie Sklodowska-Curie grant agreement No 702410.


\end{document}